\pdfoutput=1

\documentclass[11pt]{article}

\PassOptionsToPackage{hyphens}{url}

\usepackage[preprint]{acl}

\usepackage{times}
\usepackage{latexsym}
\usepackage[T1]{fontenc}
\usepackage[utf8]{inputenc}

\usepackage{microtype}
\usepackage{inconsolata}

\usepackage{amsmath,amssymb,amsfonts}
\usepackage{graphicx}
\usepackage{booktabs}
\usepackage{multirow}
\usepackage{makecell}
\usepackage{array}
\usepackage{tabularx}
\usepackage{subcaption}
\usepackage{tikz}
\usetikzlibrary{arrows.meta, positioning, shapes.geometric, fit, backgrounds, calc}

\usepackage{xcolor}
\definecolor{codebg}{RGB}{245, 245, 245}
\definecolor{codeframe}{RGB}{200, 200, 200}
\definecolor{trainbg}{RGB}{232, 245, 253}
\definecolor{trainedge}{RGB}{25, 118, 210}
\definecolor{infbg}{RGB}{255, 243, 224}
\definecolor{infedge}{RGB}{230, 81, 0}

\usepackage{listings}
\usepackage{enumitem}

\title{CodeGrep: An RL-Trained Retrieval Agent for LLM Coding Agents}

\author{Wuya Chen$^{1,*}$ and Yihao Yang$^{2}$ and Yang Cao$^{2}$ and Yue Lin$^{1}$ \\
  \texttt{\{chenwuya@corp.netease.com, 13590078189@163.com\}} \\
  \texttt{\{yihao786, cy1835208239\}@gmail.com} \\
  \texttt{gzlinyue@corp.netease.com} \\
  $^{1}$Netease Guangzhou AI Lab \quad
  $^{2}$Independent Researcher}

\begin{document}
\maketitle

\begin{abstract}
Modern LLM coding agents---Claude Code, OpenHands---share
a common inefficiency: they spend most of their token budget
\emph{finding} the file to patch, not patching it. On SWE-Bench
Verified, a $30$B OpenHands agent averages $23$ rounds and $631$K
tokens per resolved issue, dominated by \texttt{grep}, \texttt{glob},
and \texttt{view\_file} calls. No open, systematically-studied
agent-style code retriever exists at scale to shorten this phase.
We introduce \textbf{CodeGrep}, a $14$B retrieval \emph{agent}
trained end-to-end with GRPO to issue multi-turn parallel
\texttt{grep}/\texttt{glob}/\texttt{read} tool calls and return
candidate files to a frozen downstream coding agent.

On all $500$ SWE-Bench Verified instances, CodeGrep delivers a
small but reproducibly positive resolve-rate lift
($+1.2$pp: $25.8\%\to 27.0\%$) alongside a much larger efficiency
gain ($-15\%$ rounds, $-19\%$ tokens on resolved instances). Across
three retrievers, downstream utility follows a
\emph{precision threshold}: BM25 (precision $0.375$) degrades the
agent, Jina ($0.445$) is neutral, and only CodeGrep ($0.677$)
crosses the threshold at which retrieval begins to buy efficiency.
To enable this study, we mine supervision from $67$K open-source
agent trajectories (CATM) and build a Git-worktree environment for
multi-turn agent RL. In our setting, applying the efficiency signal
at the advantage layer, rather than the reward layer, holds KL
drift to one-third and translates cleanly into downstream
efficiency. We will release the model, training pipeline, RL
environment, and evaluation harnesses.
\end{abstract}

\section{Introduction}
\label{sec:intro}

Modern LLM coding agents---general-purpose assistants like
Claude Code~\cite{claudecode} and open GitHub-issue-solving
frameworks like OpenHands~\cite{openhands}---share a common
inefficiency in how they allocate their token budget: much of it is
spent \emph{finding} the file to patch, not writing the patch
itself. In
this work, we ground this observation on a concrete instance: a
$30$B OpenHands agent on SWE-Bench Verified~\cite{swebench}
averages $23$ rounds and $631$K tokens per resolved issue, with a
large fraction consumed by \texttt{grep}, \texttt{glob}, and
\texttt{view\_file} calls issued in ambiguous exploration loops---often
$20{+}$ rounds in a wrong direction before the agent gives up.

Can a specialized retrieval submodule shorten this exploration
phase? Two hypotheses have not been jointly tested.
\textbf{H1 (effectiveness)}: better file-level retrieval steers the
agent past dead-end explorations and raises the resolve rate.
\textbf{H2 (efficiency)}: the agent resolves roughly the same set
of issues but with a shorter rollout. Distinguishing them requires
evaluating a retriever both on its intrinsic quality and on its
downstream effect within the same agent stack---a comparison
absent from prior open code retrievers, from lexical
(BM25~\cite{bm25}) to static dense
encoders~\cite{codebert, graphcodebert, jinacodeembed}.

We train \textbf{CodeGrep}, a $14$B retrieval agent optimized
end-to-end with GRPO~\cite{grpo}, and evaluate it against BM25 and
Jina on SWE-Bench Verified. \emph{Neither H1 nor H2 holds
uniformly}: retrieval quality translates downstream through a
\emph{precision threshold}, not a linear payoff. Below the
threshold, BM25 (precision $0.375$) degrades the agent; near it,
Jina ($0.445$) is neutral; above it, CodeGrep ($0.677$) cuts $15\%$
rounds and $19\%$ tokens on resolved instances, while adding a
small but reproducibly positive $+1.2$pp to the resolve rate.

Three engineering ingredients enable this study. First,
\textbf{CATM} (Code Agent Trajectory Mining) mines relevance labels
from $67$K open-source OpenHands trajectories without human
annotation (\S\ref{sec:catm}). Second, a Docker-free Git-worktree
sandbox reduces per-rollout environment setup from minutes to
milliseconds, making multi-turn agent RL tractable on a single
$8{\times}$B200 node (\S\ref{sec:env}). Third, a systematic reward
study across three trained iterations
($v_1{\to}v_2{\to}v_3$) yields two observations about GRPO reward
design: in our setting, applying an efficiency signal at the
advantage layer is markedly more stable than at the reward layer,
and removing an auxiliary component absent from the downstream
tool improves both training stability and downstream
efficiency (\S\ref{sec:reward}).

\paragraph{Contributions.}
\begin{itemize}[leftmargin=1.5em, itemsep=2pt, topsep=3pt]
  \item \textbf{CodeGrep}, an open $14$B agent-style code retriever
        trained end-to-end with GRPO; injected into a frozen
        OpenHands agent, it cuts rounds by $15\%$ and tokens by
        $19\%$ on resolved SWE-Bench Verified instances while
        lifting resolve rate by $+1.2$pp.
  \item \textbf{A \emph{precision-threshold} characterisation of
        retrieval--agent coupling}: across three retrievers
        (BM25 $0.375$, Jina $0.445$, CodeGrep $0.677$), downstream
        utility exhibits three regimes---hurts, neutral, buys
        efficiency (\S\ref{sec:precision-threshold}).
  \item \textbf{A reward-design study} showing that (i) reward-layer
        multiplicative scaling of a tool-call penalty triples GRPO
        policy drift vs.\ advantage-layer scaling (KL $0.31$ vs.\
        $0.09$); (ii) in our final iteration, dropping an auxiliary
        line-range component from the base reward improves both
        training stability and downstream efficiency
        (\S\ref{sec:reward}).
  \item \textbf{Full open release}: model, CATM pipeline, RL
        environment, and all evaluation harnesses.
\end{itemize}

\section{Related Work}
\label{sec:related}

\paragraph{Code retrieval for SWE-Bench-style agents.}
Static retrievers---dense code
encoders~\cite{codebert, graphcodebert, unixcoder, jinacodeembed}
and lexical baselines such as BM25~\cite{bm25}---struggle when
issue vocabulary does not overlap with the target code, a common
failure mode in GitHub bug reports (\S\ref{sec:exp-retrieval}).
Pipeline systems such as Agentless~\cite{agentless},
Moatless~\cite{moatless}, and SWE-Fixer~\cite{swefixer} pair
lightweight retrieval with LLM-based reranking, but keep the
retrieval component non-trainable.

\paragraph{Agent-style search and tool use.}
Building on ReAct~\cite{react} and Toolformer~\cite{toolformer},
coding agents such as OpenHands~\cite{openhands} and
SWE-agent~\cite{sweagent} operate over rich tool interfaces at the
cost of dozens of sequential calls per issue. For search
specifically, Search-R1~\cite{searchr1} and
ReasonIR~\cite{reasonir} train LLMs to use search engines with RL,
but target indexed web corpora rather than repository file systems.

\paragraph{Reinforcement learning for tool policies.}
GRPO~\cite{grpo} has been adapted to multi-turn tool use in
ToRL~\cite{torl}, RAGEN~\cite{ragen}, and concurrent
Cognition SWE-grep~\cite{cognitionsweGrep}. A recurring
difficulty is reward-length exploitation when the reward mixes
task accuracy with efficiency signals. Our
$v_1{\to}v_2{\to}v_3$ study (\S\ref{sec:reward}) traces one
manifestation and its resolution: moving the efficiency signal from
the reward to the advantage layer.

\paragraph{Trajectory mining as supervision.}
Learning ``what is relevant'' from past agent reads is behavioural
supervision with a long history in classical IR. LRAT~\cite{lrat}
applied this to web-search agents with one-to-one browse-relevance
pairs. CATM (\S\ref{sec:catm}) adapts the template to code
retrievers that emit multiple parallel reads per turn, requiring
one-to-many attribution and per-trajectory aggregation
(Table~\ref{tab:catm-vs-lrat}); our source data are open-source
OpenHands trajectories~\cite{nebiusopenhands}.

\section{Method}
\label{sec:method}

\subsection{System Overview}
\label{sec:overview}

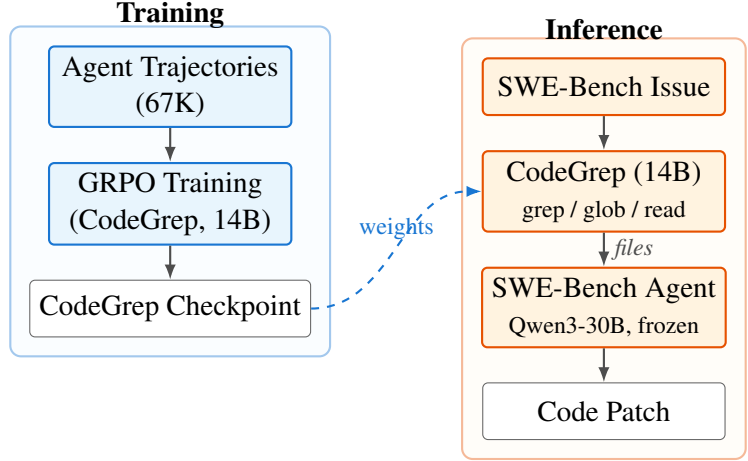
\begin{figure*}[t]
\centering
\begin{tikzpicture}[
    node distance=0.45cm and 1.0cm,
    box/.style={rectangle, rounded corners=2pt, draw, minimum width=3.2cm,
                minimum height=0.75cm, align=center, font=\normalsize},
    trainbox/.style={box, fill=trainbg, draw=trainedge, thick},
    infbox/.style={box, fill=infbg, draw=infedge, thick},
    outbox/.style={box, fill=white, draw=black!60},
    arr/.style={-{Latex[length=2mm]}, thick, black!70},
    label/.style={font=\small\itshape, black!60},
    stage/.style={font=\bfseries\normalsize},
]

\node[trainbox] (traj)  {Agent Trajectories\\(67K)};
\node[trainbox, below=of traj] (grpo)   {GRPO Training\\(CodeGrep, 14B)};
\node[outbox,   below=of grpo] (ckpt)  {CodeGrep Checkpoint};

\draw[arr] (traj)  -- (grpo);
\draw[arr] (grpo)  -- (ckpt);

\node[stage, above=0.1cm of traj] {Training};

\node[infbox, right=2.5cm of traj]     (iss)   {SWE-Bench Issue};
\node[infbox, below=of iss]     (grep)  {CodeGrep (14B)\\{\small grep / glob / read}};
\node[infbox, below=of grep]    (agent) {SWE-Bench Agent\\{\small Qwen3-30B, frozen}};
\node[outbox, below=of agent]   (patch) {Code Patch};

\draw[arr] (iss)   -- (grep);
\draw[arr] (grep)  -- node[right, font=\small\itshape, black!70] {files} (agent);
\draw[arr] (agent) -- (patch);

\node[stage, above=0.1cm of iss] {Inference};

\draw[arr, dashed, thick, trainedge]
    (ckpt.east) to[out=0, in=180] node[above, font=\small, trainedge] {weights} (grep.west);

\begin{scope}[on background layer]
    \node[fit=(traj)(grpo)(ckpt), draw=trainedge!40, thick, rounded corners=4pt,
          fill=trainbg!20, inner sep=0.25cm] {};
    \node[fit=(iss)(grep)(agent)(patch), draw=infedge!40, thick, rounded corners=4pt,
          fill=infbg!20, inner sep=0.25cm] {};
\end{scope}

\end{tikzpicture}
\caption{System overview. Training (blue) produces CodeGrep weights
from open-source agent trajectories via GRPO. Inference (orange)
uses the trained CodeGrep as a retrieval submodule whose output is
injected into a \emph{frozen} downstream agent's prompt.}
\label{fig:overview}
\end{figure*}

The system runs in two decoupled phases
(Fig.~\ref{fig:overview}). During \emph{training}, CodeGrep is
optimized with GRPO in a custom RL environment
(\S\ref{sec:env}) against relevance labels mined by CATM
(\S\ref{sec:catm}). During \emph{inference}, the trained CodeGrep
takes the issue description, emits a short list of candidate files,
and those files are injected into the prompt of a frozen OpenHands
downstream agent. Only the retriever is trained; the downstream
agent is untouched, so any downstream measurement
(\S\ref{sec:exp}) is unambiguously attributable to what the
retriever contributes to the prompt.

\subsection{CodeGrep: The Retrieval Model}
\label{sec:model}

CodeGrep is a $14$B-parameter retriever built on
Qwen3-14B-Instruct~\cite{qwen3}. It exposes three read-only
primitives---\texttt{grep} (regex search), \texttt{glob} (path
matching), and \texttt{read} (file contents)---mirroring how a
developer navigates an unfamiliar codebase. At each turn it emits
up to $8$ tool calls executed concurrently; observations are
appended to the context, and exploration terminates on a final
answer or after $4$ turns ($3$ exploration + $1$ answer,
$24$ effective reads). The answer conforms to a fixed JSON schema:
\begin{center}
\small\texttt{<answer>\{"files": [\ldots],}\\
\small\texttt{"line\_ranges": [\ldots]\}</answer>}
\end{center}
the \texttt{line\_ranges} field is dead information downstream (the
OpenHands editing tool consumes only file paths), so we drop it
from the training objective in $v_3$ (\S\ref{sec:reward}).

\subsection{RL Environment: A Lightweight Multi-turn Tool Sandbox}
\label{sec:env}

An RL environment for CodeGrep must expose the target repository at
a specific commit so that \texttt{grep}/\texttt{glob}/\texttt{read}
calls can execute and their observations feed back into the rollout.
The default of reusing official SWE-Bench Docker images is
intractable at single-node training scale: each image is $1$--$3$ GB,
pulls take minutes, and thousands of them need terabyte-scale disk.
Because CodeGrep training only reads code, the Python environment
inside each image is dead weight.

We replace Docker with a \textbf{worktree-based sandbox} whose
per-rollout setup completes in milliseconds. From any
SWE-Bench-style dataset we extract unique \emph{(repo, commit)}
pairs; for each repo we keep a single \texttt{git clone --bare} and
use \texttt{git worktree add} to instantiate one lightweight tree
per commit (disk drops from $N\cdot\text{repo\_size}$ to
$\text{repo\_size} + N\cdot\text{worktree\_size}$, with no network
I/O). At rollout time,
\texttt{grep}/\texttt{glob}/\texttt{read} run as native subprocesses
inside the target worktree, under path-traversal protection, a
$64$-way concurrency limit, and per-tool timeouts. Environment
interaction per rollout drops from \emph{minutes} to
\emph{milliseconds}, making the training runs in
\S\ref{sec:recipe} feasible on a single $8{\times}$B200 node. Full
plumbing, rollout scheduler, and the three-layer architecture
diagram are in Appendix~\ref{app:infra}.

\subsection{Training Data: CATM (Code Agent Trajectory Mining)}
\label{sec:catm}

\paragraph{Ground-truth problem and fix.}
The natural ``relevant files'' label---files touched by the
SWE-Bench gold patch---is incomplete: it omits auxiliary files that
must be \emph{read} to understand the fix but are not themselves
edited. Directly optimizing against the patch signal therefore
biases the retriever \emph{away} from the behaviour we want. We
instead adopt a behavioural notion of relevance: a file is relevant
if some past agent, while solving the issue, opened it and produced
non-trivial reasoning grounded in its contents. \textbf{CATM} (Code
Agent Trajectory Mining) instantiates this notion for code
retrieval and for agents that emit parallel tool calls; a related
notion has been used by LRAT~\cite{lrat} in the web-search setting
(see Appendix Table~\ref{tab:catm-vs-lrat} for a detailed
comparison).

\paragraph{Three-stage pipeline.}
CATM operates on $67{,}074$ open-source OpenHands
trajectories \cite{nebiusopenhands}.
\emph{Stage 1 (mining):} extract every file-read tool call
(\texttt{str\_replace\_editor} with \texttt{view}), normalize paths,
and discard directories, documentation
(\texttt{README.md}, \texttt{issue.md}), and extensionless files.
For each surviving file $f$ record the following assistant message
as the \emph{post-reasoning}, with token length $l(f)$.
\emph{Stage 2 (judge filtering):} an LLM judge classifies each
file's post-reasoning as \texttt{RELEVANT} or
\texttt{NOT\_RELEVANT}; we adopt a conservative bias treating any
non-explicit-\texttt{NOT\_RELEVANT} output as \texttt{RELEVANT}.
\emph{Stage 3 (intensity-aware weighting):} following
LRAT~\cite{lrat}, each file receives an exponential-saturation
score of its reasoning length,
\begin{align}
\tilde{w}(f) &= 1 - \exp\bigl(-\ln 2 \cdot l(f)/\beta\bigr), \\
w(f) &= \tilde{w}(f)/\mu_{\text{raw}},
\label{eq:weight}
\end{align}
where $\beta$ is the median $l(f)$ and $\mu_{\text{raw}}$ is the mean
of $\tilde{w}$; the normalization gives $\mathbb{E}[w]\approx 1$ so a
single threshold generalizes across issues.

\paragraph{Ground truth structure.}
\label{sec:catm-merge}
CATM-mined files enter the training signal not as an auxiliary
recall term but through the ground-truth set itself. For each issue
$x$, the reward's target set is
$\mathcal{G}(x) = \mathcal{G}_{\text{patch}}(x) \cup
\{f \in \mathcal{L}(x) : w_f \geq 0.15\}$,
where $\mathcal{G}_{\text{patch}}(x)$ are SWE-Bench gold-patch files
and $\mathcal{L}(x)$ is the CATM-mined set. The weights act as a
noise filter; every surviving file is a hard positive in the
$F_\beta$ reward, on equal footing with patch files. Line ranges
from the patch are used as an auxiliary scoring input for
$v_1$/$v_2$ and dropped in $v_3$ (\S\ref{sec:reward}). The
pipeline yields $31{,}977$ effective training samples ($47.7\%$
retention); the discarded remainder is dominated by misdirected
reads and documentation lookups, confirming that judge filtering
is necessary.

\subsection{Reward Design and Three Iterations}
\label{sec:reward}

Let $x$ be an issue and $\mathcal{P}(x)$ the file set predicted by a
rollout. The target $\mathcal{G}(x)$ is the CATM merge
(\S\ref{sec:catm}).

\paragraph{Component scores.}
Two precision-biased $F_\beta$ statistics ($\beta = 0.5$) appear
across the iterations. \emph{File-level}
$F_\beta^{\text{file}}(\mathcal{P}_f, \mathcal{G}_f)$ is a standard
$F_\beta$ over predicted and target files (with the natural
degenerate cases: $1$ if both are empty, $0$ if exactly one is
empty). The precision bias mirrors the downstream constraint: false
positives pollute the agent's context
(§\ref{sec:exp-downstream}), while a miss is usually recoverable.
\emph{Line-range} $F_\beta^{\text{lr}}(\mathcal{P}_r,
\mathcal{G}_r)$ is defined analogously, but matching predicted
against gold ranges: a prediction $p$ matches a gold $g$ iff they
share a filename and $p$ covers at least $50\%$ of $g$; matching is
greedy and one-to-one. Under merge mode, predicted ranges for
CATM-only files are dropped before scoring. The $F_\beta$ formula,
edge cases, and matching algorithm are in
Appendix~\ref{app:reward}.

\textbf{Rollout-level efficiency signal.}
Let $C_{\text{total}}$ be the total tool calls a rollout issues and
$T$ its number of turns. The \emph{average tool calls per turn} is
\begin{equation}
\bar{c} \;:=\; C_{\text{total}}/T \quad (T{\ge}1,\ 0{\le}\bar{c}{\le}8),
\label{eq:cbar}
\end{equation}
with the $8$ from the per-turn parallel-call budget
(\S\ref{sec:model}). We treat $\bar{c}=4$ as the natural saturation
point.

\paragraph{Iteration $v_1$: reward-layer efficiency scaling.}
Average the two component scores and multiply by a reward-level
efficiency scale:
\begin{equation}
\boxed{\ R^{v_1}
= \tfrac{1}{2}\bigl(F_\beta^{\text{file}} + F_\beta^{\text{lr}}\bigr)
\cdot \sigma^{v_1}(\bar{c})\ }
\end{equation}
\begin{equation}
\sigma^{v_1}(\bar{c})
= \frac{1}{\max(1,\;\bar{c}/4)} \in (0, 1].
\end{equation}
$\sigma^{v_1}$ is piecewise: no penalty for $\bar{c}\le 4$, a
$1/(\bar{c}/4)$ discount above. Trained $v_1$ checkpoints exhibit
two coupled failure modes: policy drift is elevated and
per-instance downstream efficiency is negligible.
Section~\ref{sec:reward-iterations} provides the numerical
diagnosis.

\paragraph{Iteration $v_2$: move scaling to the advantage layer.}
Keep the base score, drop the reward-level scale to a hard mask on
degenerate rollouts, and move the discount into the GRPO advantage
estimator:
\begin{equation}
\boxed{\ R^{v_2}
= \tfrac{1}{2}\bigl(F_\beta^{\text{file}} + F_\beta^{\text{lr}}\bigr)
\cdot \mathbf{1}_{\bar{c}>0}\ }
\end{equation}
\vspace{-1em}
\begin{equation}
\boxed{\ A^{v_2}_i = A_i \cdot s(\bar{c}_i)\ }, \quad
s(\bar{c}) = \sqrt{\min(\bar{c}/4, 1)}.
\label{eq:adv-scale}
\end{equation}
The square root softens the discount: rather than penalizing
low-$\bar{c}$ rollouts linearly, $s(\bar{c})$ preserves a
non-trivial fraction of the gradient signal even at $\bar{c}=1$
($s=0.50$), while flattening once $\bar{c}$ passes the half-budget
saturation point $\bar{c}=4$ ($s=1$). Two properties motivate the
split:
(i) the group ranking of raw rewards is preserved so GRPO's
intra-group comparison stays anchored to task performance;
(ii) the discount applies at the gradient step for a specific
rollout, not to other group members' rewards. $v_2$ resolves the
drift symptom but develops a new one (length exploitation), taken up
in \S\ref{sec:reward-iterations}.

\paragraph{Iteration $v_3$: remove line-range from the base reward.}
Drop the line-range component entirely:
\begin{equation}
\boxed{\ R^{v_3} = F_\beta^{\text{file}}(\mathcal{P}_f, \mathcal{G}_f)
\cdot \mathbf{1}_{\bar{c}>0}\ }
\end{equation}
with the same $s(\bar{c})$ advantage scaling
(Eq.~\ref{eq:adv-scale}). Rationale:
(i) \emph{Interface mismatch:} the downstream editor has no
\texttt{view\_range} argument, so predicted line ranges are dead
information at inference time.
(ii) \emph{Systematic label bias:} CATM-only entries have no gold
line ranges by construction, so $F_\beta^{\text{lr}} = 0$ for a large
fraction of training data.

The concurrent Cognition system~\cite{cognitionsweGrep} retains a
line-range term; a component-level comparison
(Appendix~\ref{app:reward}, Table~\ref{tab:reward-vs-blog}) is
necessarily partial as their reward formulation is not fully
disclosed. Table~\ref{tab:reward-versions} summarizes our three
iterations.

\begin{table}[t]
\centering
\small
\renewcommand{\arraystretch}{1.25}
\setlength{\tabcolsep}{4pt}
\caption{Summary of the three reward-design iterations. All variants
share the same target $\mathcal{G}$ and $\bar{c}$ signal
(Eq.~\ref{eq:cbar}).}
\label{tab:reward-versions}
\begin{tabular}{p{2.6cm}ccc}
\toprule
\textbf{Component} & \textbf{$v_1$} & \textbf{$v_2$} & \textbf{$v_3$} \\
\midrule
Base task score
    & $\tfrac{1}{2}(F_\beta^{\text{file}}{+}F_\beta^{\text{lr}})$
    & $=v_1$
    & $\mathbf{F_\beta^{\text{file}}}$ \\
Reward-layer scale
    & $\frac{1}{\max(1, \bar{c}/4)}$
    & $\mathbf{1}_{\bar{c}>0}$
    & $\mathbf{1}_{\bar{c}>0}$ \\
Advantage-layer scale
    & ---
    & $s(\bar{c})$
    & $=v_2$ \\
\bottomrule
\end{tabular}
\end{table}

\subsection{Training Recipe}
\label{sec:recipe}

We train CodeGrep with GRPO on top of ms-swift~\cite{msswift} on a
single $8{\times}$B200 node, using LoRA~\cite{lora} plus optimizer
offload and vLLM-colocated rollout to fit the $14$B run on one node.
End-to-end wall-clock is $\sim\!27$ hours per run; we early-stop at
step $897$ of a planned $2000$-step budget. Full hyperparameters
(LoRA rank, batch/rollout config, LR schedule, sampling) are in
Appendix~\ref{app:training-config}.

\begin{figure*}[t]
\centering
\includegraphics[width=\textwidth]{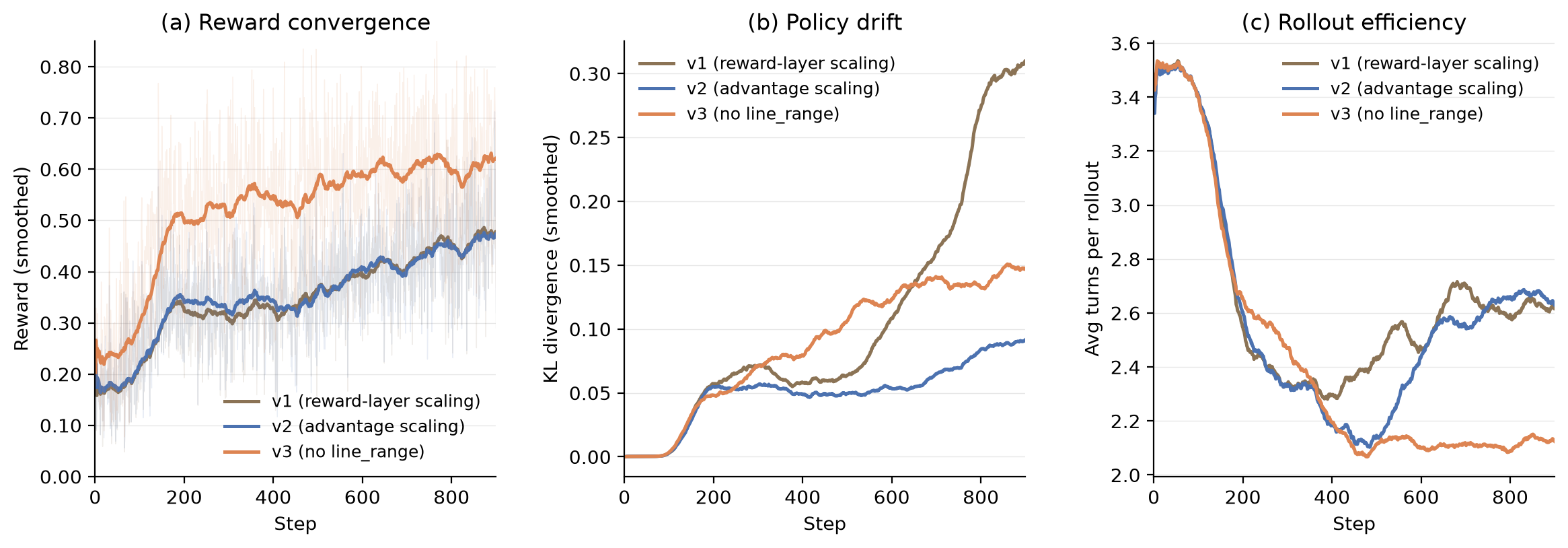}
\caption{Training dynamics of the three reward-design iterations.
\textbf{(a)} Reward: $v_3$ climbs faster and levels off higher;
$v_1$ and $v_2$ plateau together.
\textbf{(b)} KL-to-reference: $v_1$'s reward-layer scaling drives KL
to $\sim\!0.31$; $v_2$/$v_3$ hold at $0.09$/$0.15$.
\textbf{(c)} Tool-use turns: $v_3$ stabilizes at $\sim\!2.1$; $v_1$
and $v_2$ rebound to $\sim\!2.6$ after step $500$.}
\label{fig:training-curves}
\end{figure*}

\paragraph{Training dynamics.}
Fig.~\ref{fig:training-curves} tracks reward, KL-to-reference, and
average rollout length across $v_1$/$v_2$/$v_3$. Three signatures
stand out: $v_3$ converges highest on reward ($0.60$--$0.65$ vs.\
$0.45$--$0.48$ for $v_1$/$v_2$); $v_1$'s KL drifts to $\sim\!0.31$
by step $900$ while $v_2$/$v_3$ stay at $0.09$/$0.15$; average
tool-use turns drop to $\sim\!2.1$ under $v_3$ and remain there,
while $v_1$/$v_2$ rebound to $\sim\!2.6$ after step $500$.
Together these are the training-side evidence for the reward-design
principles unpacked in \S\ref{sec:reward-iterations}. Additional
diagnostics (completion length, gradient norm, clipping ratio) are
in Appendix~\ref{app:stability}.

\section{Experiments}
\label{sec:exp}

\subsection{Setup}
\label{sec:setup}

We evaluate on SWE-Bench Verified~\cite{swebench}
($500$ instances), using OpenHands~\cite{openhands} with
Qwen3-30B-A3B-Instruct-2507~\cite{qwen3} as the downstream agent
(temperature $0$, $100$-round max). We report \emph{resolve rate}
plus efficiency measures (average rounds and total tokens on
resolved instances; average rounds on unresolved). Without any
retrieval, this baseline resolves $25.8\%$ of instances at $23.0$
rounds and $631$K tokens per resolved issue---within one standard
deviation of the publicly reported $25.2\pm 0.7\%$ for the same
configuration~\cite{nebiusopenhands}, reproducing the reference
number all subsequent comparisons anchor to. We compare six
configurations: baseline, BM25 (Appendix~\ref{app:bm25}),
Jina-1.5B~\cite{jinacodeembed} (dense embedding retriever),
and the three CodeGrep iterations $v_1$/$v_2$/$v_3$. Each retriever
is evaluated first on its own retrieval quality
(\S\ref{sec:exp-retrieval}), then on downstream injection
(\S\ref{sec:exp-downstream}).

\subsection{Retrieval evaluation}
\label{sec:exp-retrieval}

The internal evaluation set is constructed from held-out
\texttt{swe-rebench} instances with CATM-mined labels, subsequently
audited by senior software engineers on our team to filter cases in
which the LLM judge over-accepts marginally-relevant reads. Each
retriever is scored with $F_\beta$ ($\beta{=}0.5$), together with
file-level precision and recall. We choose $\beta{<}1$ because a
false positive (an irrelevant file injected into the downstream
agent) is more costly than a false negative: the former inflates
the agent's context and dilutes attention with distractor code,
whereas the latter can often be recovered by the agent's own
tool calls.
Table~\ref{tab:retrieval-full} consolidates the five retrievers.

\begin{table*}[t]
\centering
\small
\renewcommand{\arraystretch}{1.25}
\setlength{\tabcolsep}{5pt}
\caption{Retrieval quality on the internal evaluation set. The
``Turns'' column is specific to the agentic retriever and does not
apply to non-agentic baselines.}
\label{tab:retrieval-full}
\begin{tabular}{lcccccc}
\toprule
\textbf{Retriever} & \textbf{$F_\beta$ mean} & \textbf{$F_\beta$ median}
& \textbf{Precision} & \textbf{Recall}
& \textbf{$F_\beta\ge 0.8$} & \textbf{Turns} \\
\midrule
BM25                          & $0.359$ & $0.455$ & $0.375$ & $0.386$ & $7.0\%$ & -- \\
Jina-1.5B                     & $0.427$ & $0.500$ & $0.445$ & $0.468$ & $7.0\%$ & -- \\
CodeGrep $v_1$                & $0.562$ & $0.556$ & $0.641$ & $\mathbf{0.486}$ & $36.7\%$ & $3.8$ \\
CodeGrep $v_2$                & $0.526$ & $0.556$ & $0.589$ & $0.483$ & $31.7\%$ & $2.9$ \\
\textbf{CodeGrep $v_3$}
                              & $\mathbf{0.576}$ & $\mathbf{0.714}$
                              & $\mathbf{0.677}$ & $0.435$
                              & $\mathbf{43.0\%}$ & $\mathbf{2.3}$ \\
\bottomrule
\end{tabular}
\end{table*}

Three observations follow.
(i) \textbf{CodeGrep dominates both BM25 and Jina, with the gap
driven primarily by precision.} $v_3$'s mean $F_\beta$ is
$1.6\times$ BM25's and $1.35\times$ Jina's, and the three retrievers
form a strict precision ordering: BM25 ($0.375$) $<$ Jina ($0.445$)
$<$ CodeGrep~$v_3$ ($0.677$). The gap is even more pronounced in the
tail: CodeGrep~$v_3$ produces high-quality retrievals
($F_\beta{\ge}0.8$) on $43.0\%$ of instances, a $6.1\times$ increase
over both BM25 and Jina (each at $7.0\%$)---the injection regime
that matters most for downstream token savings.
(ii) \textbf{Among CodeGrep variants, $v_3$ Pareto-dominates on both
quality and cost.} Median $F_\beta$ improves to $0.714$ ($+28\%$
over $v_1$ and $v_2$), the $F_\beta{\ge}0.8$ rate reaches $43.0\%$
($+36\%$ relative to $v_2$), and the mean turn count decreases
monotonically along the iteration sequence $3.8\to 2.9\to 2.3$.
(iii) \textbf{$v_1$ and $v_2$ are quality-comparable in aggregate.}
Their median $F_\beta$ coincide at $0.556$ and their recall values
differ by less than $0.003$; the mean $F_\beta$ gap ($0.562$ vs.
$0.526$) is consistent with $v_2$'s advantage-layer efficiency
signal (\S\ref{sec:reward-iterations}) trading a small amount of
retrieval quality for a substantially more stable training run
(\S\ref{sec:reward}). The line-range component does not impede file
localization; rather, it consumes optimization capacity along a
direction the downstream agent never consumes
(\S\ref{sec:reward-iterations}).

\subsection{Downstream evaluation}
\label{sec:exp-downstream}

We now inject each retriever's output into the frozen OpenHands
agent and evaluate end-to-end.
Table~\ref{tab:full-comparison} consolidates the six configurations.

\begin{table*}[t]
\centering
\small
\renewcommand{\arraystretch}{1.25}
\setlength{\tabcolsep}{6pt}
\caption{Downstream comparison of the baseline, BM25 top-$2$,
Jina-1.5B top-$2$, and CodeGrep $v_1$/$v_2$/$v_3$ on SWE-Bench Verified.
Efficiency columns (\emph{Resolved r.}, \emph{Resolved tok.},
\emph{Unresolved r.}) are averaged over each configuration's own
resolved / unresolved set.}
\label{tab:full-comparison}
\begin{tabular}{lcccc}
\toprule
\textbf{Config} & \textbf{Resolve} & \textbf{Resolved r.} &
\textbf{Resolved tok.} & \textbf{Unresolved r.} \\
\midrule
Baseline                    & $25.8\%$ & $23.0$ & $631$K & $32.0$ \\
BM25                        & $25.2\%$ & $22.9$ & $763$K & $29.7$ \\
Jina-1.5B                   & $25.8\%$ & $23.2$ & $587$K & $27.8$ \\
CodeGrep $v_1$              & $\mathbf{27.0\%}$ & $22.7$ & $627$K & $26.2$ \\
CodeGrep $v_2$              & $26.6\%$ & $21.4$ & $584$K & $26.4$ \\
\textbf{CodeGrep $v_3$}     & $\mathbf{27.0\%}$ & $\mathbf{19.6}$ &
$\mathbf{514}$K & $\mathbf{27.5}$ \\
\bottomrule
\end{tabular}
\end{table*}

Four observations follow.
(i) \textbf{$v_3$ delivers a small but reproducibly positive
resolve-rate lift alongside a much larger efficiency dividend.}
Resolve rate lifts $+1.2$pp ($25.8\%\to 27.0\%$), while resolved
rounds drop $23.0\to 19.6$ ($-15\%$) and resolved tokens
$631$K$\to 514$K ($-19\%$).
(ii) \textbf{BM25 injection \emph{degrades} downstream.} Resolve
drops $0.6$pp and resolved tokens inflate $21\%$ ($631$K$\to 763$K);
within the $94$ instances both configs resolve, BM25 rollouts spend
$6.6\%$ more rounds and $38.6\%$ more tokens.
(iii) \textbf{Jina injection is roughly neutral.} Resolve rate is
identical to baseline ($25.8\%$) and resolved tokens drop $7\%$
($631$K$\to 587$K)---a modest efficiency shift with no resolve
lift, placing Jina between BM25 and CodeGrep.
(iv) \textbf{$v_1$ matches $v_3$'s resolve rate but not its
efficiency.} $v_1$ hits $27.0\%$ resolve but only $-1.3\%$ rounds
and $-0.6\%$ tokens---essentially flat; $v_2$ sits between, reaching
$26.6\%$ resolve with $-7\%$ rounds and $-7\%$ tokens.

\section{Analysis}
\label{sec:analysis}

We now unpack two non-obvious patterns visible across
Tables~\ref{tab:retrieval-full}--\ref{tab:full-comparison}:
(i) how retrieval quality translates to downstream utility via a
precision threshold, and (ii) how the three reward iterations
induce three qualitatively distinct downstream regimes. A
methodological caveat on efficiency accounting follows in
\S\ref{sec:pooled-vs-paired}.

\subsection{From retrieval quality to downstream utility: a precision threshold}
\label{sec:precision-threshold}

A joint reading of Tables~\ref{tab:retrieval-full} and
\ref{tab:full-comparison} reveals that retrieval quality does not
translate linearly into downstream utility. Instead, we observe a
monotone precision gradient partitioned into three regimes.

(i) \textbf{Below threshold, retrieval hurts.} BM25, at
file-precision $0.375$, lowers resolve rate by $0.6$pp and inflates
resolved tokens by $21\%$ relative to baseline; restricted to the
$94$ instances both configurations resolve, BM25 rollouts spend
$6.6\%$ more rounds and $38.6\%$ more tokens. Low-precision
candidates function as distractors from which the agent must
recover.

(ii) \textbf{Near threshold, injection is roughly neutral.} A dense
embedding retriever (Jina-1.5B) at precision $0.445$ leaves the
resolve rate at baseline ($25.8\%$) and yields only a modest
efficiency shift ($-7\%$ resolved tokens). The direction of change
is consistent with CodeGrep, but the magnitude is an order of
magnitude smaller: a moderately better retriever than BM25 is
insufficient to unlock a downstream dividend.

(iii) \textbf{Above threshold, marginal quality gains are absorbed
by rollout efficiency.} Both CodeGrep variants already lift resolve
rate above baseline (\S\ref{sec:exp-downstream}); comparing further
along the CodeGrep axis, $v_2\to v_3$ improves median $F_\beta$ by
$+28\%$ ($0.556\to 0.714$) and the $F_\beta{\ge}0.8$ rate by
$+36\%$, yet the resolve rate advances by only $+0.4$pp while
resolved rounds and tokens fall $8\%$ and $12\%$ respectively. Once
the retriever is precise enough to be net-useful, additional
precision no longer expands the resolvable set; it is absorbed as
rollout compression.

Two mechanisms are consistent with this gradient:
(a) file localization is not the binding bottleneck on
resolve rate---the downstream agent already recovers much of the
task-relevant context through its own \texttt{grep}/\texttt{view}
tools on the instances it can resolve, and on the instances it
cannot, the failure is typically in patch synthesis rather than in
locating the right files; sharper retrieval therefore compresses
rollouts rather than enlarging the resolvable set;
(b) the marginal harm of a false positive is comparable in
magnitude to the marginal benefit of a true positive, so a
low-precision retriever can be net-negative in aggregate.

\textbf{Takeaway.} Retrieval quality translates to downstream
performance through a \emph{precision threshold} rather than a
linear payoff. The three data points---BM25 ($0.375$, hurts),
Jina ($0.445$, neutral), and CodeGrep ($0.677$, buys
efficiency)---trace a monotone gradient, placing the transition
from neutral to net-positive somewhere between precision $0.45$ and
$0.68$. Whether the exact threshold transfers to other downstream
agents remains open (\S\ref{sec:limitations}).

\subsection{Reward iterations: three failure modes, three regimes}
\label{sec:reward-iterations}

We now unpack the failure modes underlying the reward transitions
introduced in \S\ref{sec:reward}. Evidence is drawn from the
training curves (Fig.~\ref{fig:training-curves},
Appendix~\ref{app:stability}) and
Tables~\ref{tab:retrieval-full}--\ref{tab:full-comparison}.

\paragraph{$v_1\to v_2$: reward-layer scaling destabilises training
and fails to translate into downstream efficiency.}
(i) \emph{Distorted advantage estimates.} GRPO advantages are
group-relative, so multiplicative scaling on the reward reshapes
intra-group variance and distorts the sign structure of the
resulting advantage estimates. The observable consequence is a
tripled policy drift: $v_1$'s KL divergence to the reference policy
reaches $\sim\!0.31$ by step $900$ (Fig.~\ref{fig:training-curves}b),
$3.4\times$ that of $v_2$ ($\sim\!0.09$).
(ii) \emph{Higher inference cost.} $v_1$ requires $3.8$ mean turns
per inference, $1.65\times$ $v_3$'s $2.3$.
(iii) \emph{Downstream efficiency does not materialise.} $v_1$
matches $v_3$'s $27.0\%$ resolve rate but achieves only $-1.3\%$
rounds and $-0.6\%$ tokens relative to baseline, versus $-15\%$
and $-19\%$ under $v_3$.

\paragraph{$v_2\to v_3$: fixing drift exposes length exploitation.}
(i) \emph{Win:} $v_2$'s final KL settles at $\sim\!0.09$
(Fig.~\ref{fig:training-curves}b), roughly one-third of $v_1$'s.
(ii) \emph{New failure mode:} mean completion length peaks near
$\sim\!2000$ tokens around steps $400$--$500$
(Fig.~\ref{fig:training-supp}a), with a clipping ratio approaching
$20\%$; advantage-layer scaling implicitly rewards long completions
whose token totals happen to score well.
(iii) \emph{No net retrieval-quality gain:} the file-level
$F_\beta$ mean under $v_2$ is $0.526$ against $0.562$ under $v_1$;
the reward-side stability improvement is expended on completion
length rather than file localization.

\paragraph{$v_3$ closes the loop.}
Removing $F_\beta^{\text{lr}}$ eliminates a signal that has no
downstream consumer (the code editor takes no line ranges) and that
is systematically zero on CATM-only entries. Under matched
hyperparameters, $v_3$ converges higher and faster on reward
(Fig.~\ref{fig:training-curves}a: $0.60$--$0.65$ vs.\
$0.45$--$0.48$), stabilises rollout turns at $\sim\!2.1$
(Fig.~\ref{fig:training-curves}c, from $\sim\!3.5$), and delivers
the headline downstream gains reported in \S\ref{sec:exp-downstream}.

\subsection{Pooled vs.\ paired-instance efficiency}
\label{sec:pooled-vs-paired}

The efficiency numbers in Table~\ref{tab:full-comparison} are
\emph{pooled} over each configuration's own resolved set
(baseline $129$, $v_3$ $135$), so they mix a genuine efficiency
effect with a selection effect. Restricting to the $96$ instances
both configurations resolve, rounds fall $20.1\to 18.3$ ($-9\%$)
and tokens $525$K$\to 448$K ($-15\%$)---preserving $60$--$80\%$ of
the pooled effect and confirming the improvement as per-issue
acceleration.

\section{Case Study: Where the Token Savings Come From}
\label{sec:case}

We complement the aggregate numbers with an action-level trace on
a representative issue where injection converts a stalled
$74$-action rollout into a $15$-action success.

\paragraph{The issue: \texttt{django-15278}.}
\label{sec:case-django}
Adding a nullable \texttt{OneToOneField} via migration crashes on
SQLite with \emph{Cannot add a UNIQUE column}: the
\texttt{OneToOneField} constructor in
\path{django/db/models/fields/related.py} sets \texttt{unique=True}
unconditionally, forcing a UNIQUE constraint that SQLite forbids on
a nullable column.

\paragraph{Baseline: $74$ actions, $3.2$M tokens, no patch.}
Without a retrieval hint, the frozen agent stalls after $74$
actions: $35$ terminal calls dominated by repo-wide \texttt{find}
and \texttt{grep}, $17$ chain-of-thought turns, and $18$
file-editor operations touching $13$ unique files, none of which
contains the fix. Its first $18$ tool calls
(Table~\ref{tab:case-django-trajectory} in
Appendix~\ref{app:case-trajectory}) expose the failure mode: the
agent searches for \texttt{oauth2}---a token from the issue's
third-party reproducer, absent from Django itself---inspects
\path{tests/migrations/test_state.py} for a related test, and
eventually attempts an edit in
\path{django/db/migrations/operations/fields.py}, which controls
\emph{how} \texttt{AddField} runs rather than \emph{why}
\texttt{OneToOneField} declares itself unique.

\paragraph{Injection: $15$ actions, $307$K tokens, patch passes.}
The injected \texttt{<retrieved\_context>} lists three candidate
files, including \path{related.py}. The same agent deliberates for
five turns, opens \path{related.py} at action $6$, greps for
\texttt{OneToOneField} at action $8$ to locate the class
definition, applies a minimal edit at action $12$ conditioning
\texttt{unique=True} on \texttt{null=False}, and calls
\texttt{finish} at action $14$. Result: one workspace file
touched, $10\times$ fewer tokens, patch passes the harness.

CodeGrep does not supply the fix---the agent still greps to locate
the exact definition. Instead, it compresses the exploratory prefix
the frozen agent otherwise burns on repository-wide
\texttt{find}/\texttt{grep} sweeps for issue-prose tokens that
never appear in the codebase. This prefix is where the token
savings originate.

\section{Limitations and Future Work}
\label{sec:limitations}

CodeGrep's contribution is concentrated on rollout cost, not
resolve rate; a natural next step is end-to-end co-training against
a downstream resolve-rate signal, testing whether retrieval can be
made to buy both efficiency and effectiveness.

\section{Conclusion}
\label{sec:conclusion}

We introduced CodeGrep, an open $14$B RL-trained agent-style code
retriever. Injected into a frozen OpenHands downstream agent, it
lifts SWE-Bench Verified resolve rate by $+1.2$pp and cuts $15\%$
of rounds and $19\%$ of tokens on resolved instances. Our study
surfaces a \emph{precision threshold} governing when retrieval
helps downstream, and---in our training setting---identifies the
advantage layer, not the reward layer, as the appropriate site for
efficiency signals in GRPO-style multi-turn training. All artifacts
(model, training pipeline, RL environment, evaluation harnesses)
will be released.

\bibliography{custom}

\appendix

\section{Infrastructure Details}
\label{app:infra}

This appendix preserves the complete engineering realization of the RL
environment (\S\ref{sec:env}) and the CATM pipeline (\S\ref{sec:catm}).

\subsection{RL Environment Implementation}

\begin{figure*}[t]
\centering
\includegraphics[width=0.95\textwidth]{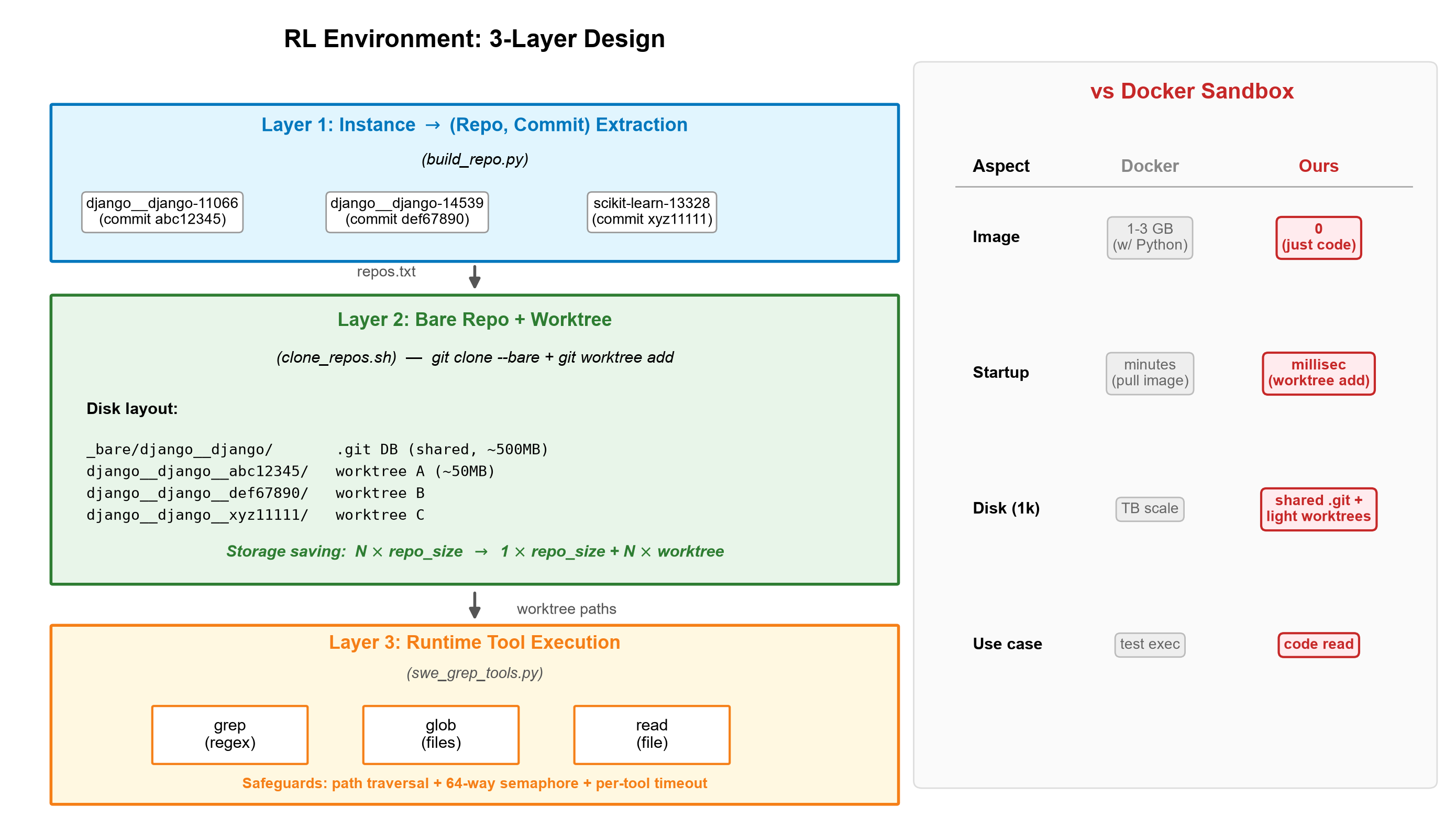}
\caption{Three-layer RL environment. Layer 1 extracts unique
(\textit{repo}, \textit{commit}) pairs; Layer 2 uses
\texttt{git clone --bare} and \texttt{git worktree add} to share the
\texttt{.git} object database across commits; Layer 3 executes
\texttt{grep}/\texttt{glob}/\texttt{read} through native subprocesses
with safety guards. The right-hand sidebar compares against a Docker
sandbox.}
\label{fig:rl-env}
\end{figure*}

A rollout begins with the model emitting up to eight parallel
\texttt{grep}/\texttt{glob}/\texttt{read} tool calls, which the sandbox
executes concurrently inside the target worktree. Their responses are
appended to the conversation with a loss mask of zero, so that only the
tokens generated by the model itself contribute to the gradient. The model
then sees the accumulated context and emits either the next round of tool
calls or a final \texttt{<answer>} block. The rollout terminates when the
model produces an answer, when no tool call is generated, when the
completion budget is exhausted, or after the configured maximum of
four turns ($3$ exploration $+ 1$ answer, matching \S\ref{sec:model}).

\paragraph{Layer 2 engineering.}
A single bare repository (e.g. \path{_bare/django__django/}) holds the
shared \texttt{.git} object database, and each target commit is
materialized as a lightweight worktree (\path{django__django__abc12345/},
\path{..._def67890/}, etc.). The initial \texttt{git clone -{}-bare} is
a one-time network operation guarded by a $600$-second timeout, and every
subsequent \texttt{git worktree add} is a local file-system operation
completing in milliseconds. To make the clone pipeline robust in the
presence of private repositories or unusual GitHub configurations, we
disable interactive prompts with \texttt{GIT\_TERMINAL\_PROMPT=0} and
\texttt{GIT\_ASKPASS=echo}, fall back from \texttt{fetch <commit>} to
\texttt{fetch -{}-all} when a commit is not directly fetchable, and
pre-validate every commit's existence with \texttt{git cat-file -e
<commit>} before invoking \texttt{worktree add}.

\paragraph{Layer 3 engineering.}
The tool executor (\texttt{swe\_grep\_tools.py}) canonicalises
every argument path and refuses any access that escapes the
repository root. A global asyncio semaphore
(\texttt{Semaphore(64)}) caps the number of concurrent subprocesses;
per-tool timeouts (10s for \texttt{grep} and \texttt{glob}, 5s for
\texttt{read}) shield the rollout from pathological queries. Tool outputs
are truncated to $4096$ characters, with $50$ matches per \texttt{grep} and
$200$ lines per \texttt{read}, so that a single call cannot exhaust the
context window. All searches use the native \texttt{grep -rn --include=*.py
...} rather than a pure-Python implementation, gaining roughly an order of
magnitude in throughput.

\paragraph{Rollout scheduler.}
\texttt{SweGrepScheduler} is a subclass of ms-swift's
\texttt{MultiTurnScheduler}. It maintains state across turns, masks tool
responses out of the loss, injects the correct worktree path per rollout,
and dispatches early termination. In total the class is roughly $400$
lines of Python.

\subsection{CATM Pipeline Implementation}

\begin{table*}[t]
\centering
\small
\renewcommand{\arraystretch}{1.35}
\setlength{\tabcolsep}{6pt}
\caption{Comparison of CATM (ours) with LRAT~\cite{lrat}.}
\label{tab:catm-vs-lrat}
\begin{tabular}{@{}p{3.3cm}p{5.3cm}p{5.3cm}@{}}
\toprule
\textbf{Dimension} & \textbf{LRAT} & \textbf{CATM (ours)} \\
\midrule
Target scenario &
Web retrieval (Wikipedia QA) &
Code retrieval (SWE-Bench) \\
Action pattern &
Sequential, $1$ browse per turn &
Parallel, up to $8$ reads per turn \\
Reasoning attribution &
One reasoning trace per document (one-to-one) &
One reasoning trace shared across $N$ parallel reads (one-to-many) \\
Sample granularity &
One (\emph{query}, \emph{doc}) pair per browse; many pairs per trajectory &
One (\emph{issue}, \emph{filtered file set}) per trajectory \\
Judge model &
Qwen3-30B-A3B-Thinking-2507 &
GLM-5.1-FP8 \\
Data scale &
$91$K pairs from $26$K trajectories &
$32$K samples from $67$K trajectories ($47.7\%$ retention) \\
Training method &
Weighted InfoNCE (dense retriever) &
GRPO (multi-turn tool-calling RL) \\
\bottomrule
\end{tabular}
\end{table*}

\paragraph{Stage 1.}
For each trajectory we traverse the message stream, identify
\texttt{role="assistant"} messages containing \texttt{tool\_calls}, and
extract \texttt{command="view"} operations along with their target paths.
We auto-detect the repository prefix
(\texttt{/workspace/owner\_\_repo\_\_commit/}) and normalize to
repo-relative paths. After a view we skip over any subsequent
\texttt{role="tool"} messages and locate the next
\texttt{role="assistant"} message, whose content is recorded as the file's
post-reasoning. If a file is viewed multiple times, we keep the instance
with the longest post-reasoning.

\paragraph{Stage 2.}
We use GLM-5.1-FP8 as the judge, prompted with the file path, the
post-reasoning, and the task description. The judge returns either
\texttt{RELEVANT} or \texttt{NOT\_RELEVANT}; when the model output is
ambiguous we default conservatively to \texttt{RELEVANT}. The judge runs
with $16$ parallel worker threads under a $60$-second per-call timeout,
and the pipeline is resumable by checkpointing processed instance IDs to
disk with real-time flushing (append mode). End-to-end the judge stage
completes in about five hours.

\paragraph{Stage 3.}
The global constants $\beta$ (the median post-reasoning length across all
mined candidates) and $\mu_{\text{raw}}$ (the mean of the unnormalized
weights over the whole dataset) require two passes over the data: the
first estimates them; the second assigns per-file weights via
Eq.~\eqref{eq:weight}. Files with $w < 0.15$ are discarded.

\subsection{Key Engineering Numbers}

Table~\ref{tab:eng-numbers} lists the salient scale parameters and
throughput figures for both the environment and the data pipeline.

\begin{table*}[t]
\centering
\small
\renewcommand{\arraystretch}{1.25}
\setlength{\tabcolsep}{6pt}
\caption{Key engineering scale and throughput.}
\label{tab:eng-numbers}
\begin{tabular}{p{6.5cm}p{7.5cm}}
\toprule
\textbf{Item} & \textbf{Value} \\
\midrule
Environment: unique (\emph{repo}, \emph{commit}) pairs &
$\sim$thousands (after dedup) \\
Environment: disk saving (bare + worktree) &
$N\!\times\!\text{repo\_size} \to \text{repo\_size} + N\!\times\!\text{worktree}$ \\
Environment: tool concurrency cap & $64$ (asyncio semaphore) \\
Environment: max parallel tools per turn & $8$ \\
Environment: max tool execs per rollout & $32$ ($4$ turns $\times$ $8$) \\
Data: raw trajectories & $67{,}074$ \\
Data: judge model & GLM-5.1-FP8 ($16$-way parallel) \\
Data: judge thresholds & reasoning\_tokens $\geq 30$, $w \geq 0.15$ \\
Data: effective training samples & $\mathbf{31{,}977}$ ($47.7\%$ retention) \\
Data: judge stage runtime & $\sim$5 hours \\
Training: total wall-clock & $27$ hours ($8\times$ B200) \\
\bottomrule
\end{tabular}
\end{table*}

\subsection{Training Configuration}
\label{app:training-config}

We train CodeGrep with GRPO on ms-swift on a single $8{\times}$B200
node.

\paragraph{Parameter-efficient fine-tuning.}
LoRA (rank $32$, $\alpha=64$) is applied to all linear projections
of the $14$B backbone; the base model weights stay frozen.
Optimizer and model states are offloaded to CPU when idle, and
training and rollout share GPUs via vLLM's colocated mode
(tensor-parallel $8$, $35\%$ of GPU memory allocated to the vLLM
engine, prefix caching on).

\paragraph{Optimization.}
Peak learning rate $5{\times}10^{-6}$ with cosine decay and $5\%$
linear warmup; GRPO KL coefficient $\beta = 0.02$. Each step
processes $64$ prompts (per-device batch $1$, gradient accumulation
$8$, across $8$ GPUs) and samples $8$ GRPO rollouts per prompt, for
$512$ effective rollouts per step. Rollouts cap at $4096$ generation
tokens and $4$ tool-use turns ($3$ exploration $+ 1$ answer);
sampling uses temperature $1.0$, top-$p$ $0.9$, top-$k$ $50$. We
early-stop at step $897$ of a planned $2000$-step budget when
reward improvement plateaus and KL-to-reference stays within safe
bounds; end-to-end wall-clock is $\sim\!27$ hours per run.

\section{Reward Design Details}
\label{app:reward}

\paragraph{Edge cases for $F_\beta^{\text{file}}$ and
$F_\beta^{\text{lr}}$.}
Both scores are precision-biased $F_\beta$ with $\beta = 0.5$,
with $\mathrm{prec}\cdot\mathrm{rec} / (\beta^2\,\mathrm{prec} +
\mathrm{rec})\cdot(1+\beta^2)$ as the general formula, plus two
degenerate cases: if both predicted and target sets are empty the
score is $1$; if exactly one is empty it is $0$.

\paragraph{Line-range matching algorithm.}
Each answer emits a list of line ranges $\mathcal{P}_r$ of the form
$(\textit{filename}, \textit{start}, \textit{end})$. The target set
$\mathcal{G}_r$ comes from the SWE-Bench gold patch (patch files
only). A predicted $p$ matches a gold $g$ iff:
(a) $p.\textit{filename} = g.\textit{filename}$; and
(b) $|[p.s, p.e]\cap[g.s, g.e]|/(g.e - g.s + 1) \ge 0.5$.
Matching is greedy over $\mathcal{P}_r$ and each side consumes at
most one match: iterating predictions in order, each pairs with the
first still-unpaired gold range it overlaps; both are then removed
from further consideration. $\mathrm{TP}$ is the number of matched
pairs; $\mathrm{prec} = \mathrm{TP}/|\mathcal{P}_r|$,
$\mathrm{rec} = \mathrm{TP}/|\mathcal{G}_r|$. Under merge mode, we
drop from $\mathcal{P}_r$ any predicted range whose filename is a
CATM-only file before scoring (such files have no gold range to
match against, and would otherwise appear as unrecoverable false
positives).

\paragraph{On merge mode and the ``no CATM recall bonus'' choice.}
In principle the pipeline supports an auxiliary CATM-recall term
with a mixing weight $\alpha\in[0,1]$ that would blend the base
task score with a soft weighted recall over CATM files. Under merge
mode this branch is inactive ($\alpha=0$), because every surviving
CATM file is already inside $\mathcal{G}$ and contributes to
$F_\beta^{\text{file}}$ directly. Keeping $\alpha=0$ makes the
three iterations cleanly comparable: they differ only in how task
and efficiency signals are combined.

\paragraph{Contrast with concurrent work.}
Cognition's concurrent commercial system~\cite{cognitionsweGrep}
describes its reward as ``an average of weighted F1 scores over
file retrieval and line retrieval tasks'' with advantages scaled
``by the average number of tool calls used per turn.'' Their blog
does not specify $\beta$, the line-range matching rule, the
tool-scale functional form, or the training data pipeline, so a
component-level comparison is partial by necessity. The most
consequential difference is our $v_3$'s removal of the line-range
term. Its outcome (higher final reward,
Fig.~\ref{fig:training-curves}a; more stable rollouts,
Appendix~\ref{app:stability}; $19\%$ downstream token reduction,
\S\ref{sec:exp-downstream}) suggests the line-range signal is not
universally beneficial. We conjecture the driver is the downstream
tool interface: agents whose editor consumes line ranges (plausibly
Windsurf's Cascade) may benefit; agents whose editor operates on
full files (OpenHands, Cursor) will not, and the extra reward
dimension becomes a training-time distractor.

\begin{table*}[t]
\centering
\small
\renewcommand{\arraystretch}{1.25}
\setlength{\tabcolsep}{5pt}
\caption{Reward-design comparison with concurrent commercial
work~\cite{cognitionsweGrep}. Their blog describes a single final
configuration; where a component's mathematical form is not
disclosed in the blog we mark it \emph{n/s} (not specified). We
list our final ($v_3$) configuration.}
\label{tab:reward-vs-blog}
\begin{tabular}{p{4.6cm}p{4cm}p{4.7cm}}
\toprule
\textbf{Reward component} &
\textbf{Cognition SWE-grep} &
\textbf{This work ($v_3$)} \\
\midrule
File-score $\beta$ (precision bias) &
Precision-biased ($\beta$ not specified) &
$\beta = 0.5$ (precision-biased) \\
Line-range scoring rule &
n/s (``F1 over line retrieval'') &
Jaccard with $\ge 50\%$ overlap $\to$ $F_{\beta=0.5}$ \\
Base score composition &
Avg.\ of file F1 and line F1 &
File-only ($F_\beta$); line-range removed \\
Training data / labels &
Proprietary; construction not disclosed &
Open unsupervised CATM pipeline (§\ref{sec:catm}) \\
Efficiency signal at reward layer &
Not described (advantage-layer scaling only) &
$v_1$: $\sigma^{v_1}(\bar c) = 1/\!\max(1,\bar c/4)$;
$v_2$/$v_3$: mask only, $\mathbf{1}_{\bar c>0}$ \\
Efficiency signal at advantage layer &
``Scale by avg tool calls per turn'' (functional form n/s) &
$s(\bar c) = \sqrt{\min(\bar c/4, 1)}$ via monkey-patch on GRPO advantage \\
Explicit reward-design study &
None reported (single final config) &
$v_1\!\to\!v_2\!\to\!v_3$, three trained runs (§\ref{sec:reward}) \\
\bottomrule
\end{tabular}
\end{table*}

\section{Positioning: Agent RL vs Traditional RLHF}
\label{app:rlhf}

For readers coming from RLHF, Table~\ref{tab:rlhf-vs-agent} highlights the
main structural differences between the setting we work in and standard
single-turn preference-based RL. This is a positioning aid rather than a
methodological claim, and is included here for context.

\begin{table*}[t]
\centering
\small
\renewcommand{\arraystretch}{1.25}
\setlength{\tabcolsep}{6pt}
\caption{Comparison of traditional RLHF with the Agent RL setting used
in this work.}
\label{tab:rlhf-vs-agent}
\begin{tabular}{p{3cm}p{5cm}p{6cm}}
\toprule
\textbf{Dimension} & \textbf{Traditional RLHF} & \textbf{Agent RL (ours)} \\
\midrule
Rollout structure & Single-turn & Multi-turn (up to $8$) \\
Action space & Single completion & Up to $8$ parallel tool calls per turn \\
Reward source & Human preference model & Environment + rule-based
$F_\beta$ \\
Ground truth & \emph{Prompt}--\emph{response} pair &
Structured multi-signal (patch + trajectory + weights) \\
Environment & None & Filesystem sandbox with real code execution \\
Loss mask & Entire completion & Tool responses must be masked out \\
Data-construction cost & Low & High (judge inference + global $\beta$
statistics) \\
\bottomrule
\end{tabular}
\end{table*}

\section{Training Stability Diagnostics}
\label{app:stability}

\begin{figure*}[t]
\centering
\includegraphics[width=\textwidth]{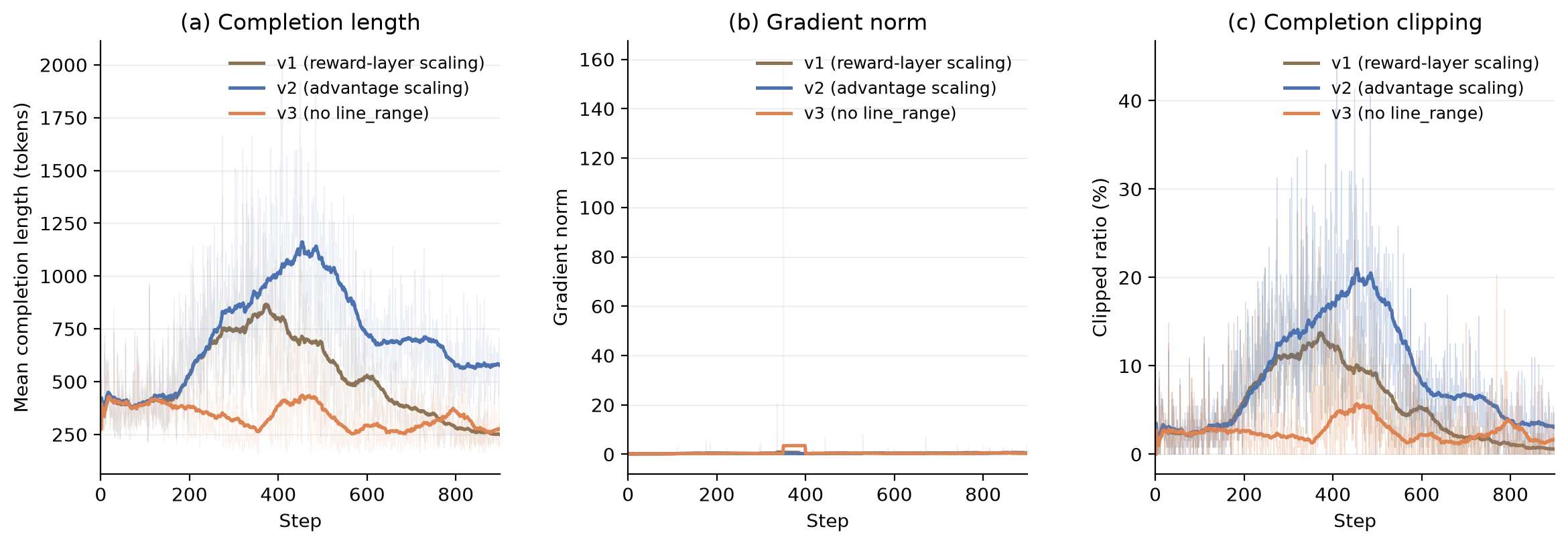}
\caption{Training-stability diagnostics over the same $\sim\!900$
GRPO steps as Figure~\ref{fig:training-curves}, for all three reward
iterations.
\textbf{(a)} Mean completion length: $v_2$ inflates dramatically
(smoothed peak $\sim\!1{,}150$ tokens near step $450$; raw peak
$\sim\!2000$), $v_1$ shows a milder mid-training bump, and $v_3$
stays near $300$--$400$ throughout.
\textbf{(b)} Gradient norms remain bounded throughout for all three.
\textbf{(c)} Completion-clipping ratio mirrors the length curve:
$v_2$ peaks at $\sim\!20\%$, $v_1$ at $\sim\!14\%$, and $v_3$ stays
below $6\%$ throughout.}
\label{fig:training-supp}
\end{figure*}

Figure~\ref{fig:training-supp} reports three additional diagnostics
tracked over the same $\sim\!900$-step training run as
Figure~\ref{fig:training-curves}: mean completion length, gradient
norm, and the fraction of clipped tokens per batch. The three panels
together characterize the training-stability regime of all three
reward configurations and complement the reward / KL / rollout
picture in the main text.

The completion-length trajectory (a) is the most revealing panel and
tells a two-story diagnostic. Under $v_2$, mean completion length
inflates from roughly $400$ tokens to a peak of $\sim\!2{,}000$ tokens
around step~$400$--$500$ before slowly retreating, and the completion
clipping ratio (c) peaks near $20\%$ during the same window---clear
evidence that $v_2$'s advantage-layer scaling implicitly rewards
\emph{relatively} short-and-well-scoring completions within a group in
a way that also permits long-completion excursions, opening a
length-exploitation channel. $v_1$ shows a milder version of the same
inflation (peak $\sim\!880$ smoothed / $\sim\!1500$ raw around step
$300$--$400$) but recovers earlier, consistent with its reward-layer
scaling directly penalising long-completion / high-turn rollouts on
the reward side. $v_3$ removes the line-range component and stays
near $\sim\!300$--$400$ throughout, with clipping ratio consistently
below $5\%$. Gradient norms (b) remain bounded throughout for all
three runs, with a single brief excursion near step~$400$ under $v_1$
and $v_2$.

Taken together, the three configurations trace a clean design
trade-off: $v_1$ (reward-layer scaling) contains the length problem
but at the cost of policy drift (KL climbing to $\sim\!0.31$;
Figure~\ref{fig:training-curves}b); $v_2$ (advantage-layer scaling)
controls policy drift but opens the length channel; only $v_3$
(no line-range) resolves both, producing training that is
simultaneously the most reward-productive, the shortest, and among
the cleanest on clipping. This picture supports our conclusion that
$v_3$'s reward is not only more effective on downstream metrics
(\S\ref{sec:exp-downstream}) but also structurally the most stable
during training.

\section{BM25 Baseline Configuration}
\label{app:bm25}

This appendix documents the BM25 baseline used for both the
retrieval-quality comparison in Table~\ref{tab:retrieval-full} and
the downstream-injection comparison in Table~\ref{tab:full-comparison},
so that both are fully reproducible.

\paragraph{Corpus.}
For each SWE-Bench-Verified instance we index the target repository's
application Python files. We skip \texttt{\_\_pycache\_\_/},
virtual-environment directories (\texttt{venv/}, \texttt{.venv/},
\texttt{env/}), and build artefacts (\texttt{build/}, \texttt{dist/},
\texttt{.eggs/}). Our default includes \texttt{tests/} and
\texttt{test/} directories in the index; the ablation below shows
that excluding them yields nearly identical results.

\paragraph{Tokenization.}
Documents and queries are tokenized identically: split on whitespace,
strip characters outside \texttt{[a-zA-Z0-9\_./]}, lower-case, drop
tokens shorter than $2$ characters, drop tokens beginning with
\texttt{http} (URL fragments in issue text). This is a deliberately
simple tokenizer intended to model a ``classical BM25'' baseline
rather than a code-specialized one.

\paragraph{BM25 parameters.}
We use BM25 Okapi with default parameters $k_1 = 1.5$ and $b = 0.75$
(the \texttt{rank\_bm25} library defaults). We did not tune these on
the evaluation set to avoid overfitting.

\paragraph{Path signal.}
We initially expected a file-path signal to help, since SWE-Bench issue
reports frequently mention directory or file-name fragments (e.g.\
``django admin form'', ``the ec2 metadata module''). We therefore built
a second BM25 index over file-path tokens and added its scores to the
content BM25 scores with a $2\times$ weight. Empirically this
augmentation had \emph{no measurable effect}: the path-augmented
variant (B) and the content-only variant (A) produce identical mean
$F_\beta$ on our evaluation set. The reason is that our tokenizer
preserves the path separators \texttt{.} and \texttt{/} within tokens,
so file-path fragments already appear as long tokens inside the content
index. We report variant B as our default for clarity, but the finding
is that a simple BM25 over tokenized code content already captures the
effect a naive path-aware BM25 could offer.

\paragraph{Top-$K$ selection.}
\label{app:bm25-topk-ablation}
We report BM25 at $K = 2$ as the default because it approximately
matches CodeGrep $v_3$'s mean output length of $1.31$ files. Larger
$K$ is not free: injecting more candidate files into the agent's prompt
inflates the per-rollout context. On retrieval quality, top-$2$
achieves $F_\beta$ mean $0.359$ (Table~\ref{tab:retrieval-full}).
For completeness we also measured $K = 5$: it raises recall (each
list is longer) at the cost of precision, and on downstream evaluation
larger $K$ carries the additional cost of longer agent-side prompts,
producing qualitatively similar downstream failure to help. We report
$K = 2$ as the main comparison because it is the fairest apples-to-apples
match to CodeGrep's output cardinality.

\paragraph{Test-directory scope.}
Our default indexes \texttt{tests/} and \texttt{test/} directories
alongside application code (via the
\texttt{-{}-include\_tests} flag), maximizing BM25's fair shot at
the ground truth (approximately $36\%$ of evaluation instances
have at least one test file in the ground truth).
Ablating this filter (i.e., skipping test directories) yields a
comparable $F_\beta$ within $0.01$, indicating that test-directory
inclusion neither meaningfully helps nor hurts BM25 on this task.
The default configuration is therefore not artificially
disadvantaging BM25.

\paragraph{Retrieval.}
At query time, we take the tokenized issue text (the user turn from
the eval prompt, which contains the bug report and error trace),
compute the combined BM25 score for every indexed file, and return the
top-$K$ by score ($K = 2$ by default). Ties are broken by insertion
order; if fewer than $K$ files have non-zero score, we still return
the top-scoring $1$--$K$ as a fallback.

\paragraph{Downstream injection.}
For the downstream evaluation (Table~\ref{tab:full-comparison}), we
inject BM25 top-$2$ predictions into the same OpenHands prompt template
used for CodeGrep. Predictions are formatted as a bulleted
\texttt{<candidate\_files>} block appended to the issue description,
byte-identical to the CodeGrep injection format. The agent then
executes its standard code-repair loop without any BM25-specific
prompting.

\paragraph{Scoring.}
Predicted files are compared against the SWE-Bench gold patch file set
using file-level $F_{\beta=0.5}$, identical to
\S\ref{sec:exp-retrieval}. We use forward-slash paths on both sides;
any residual path-separator differences are normalized before
set-intersection.

\paragraph{Runtime.}
On CPU, indexing and querying the full evaluation set completes in
under $10$ minutes on a single machine. Building the index dominates
cost; the retrieval step is under $10$~ms per query. The BM25
downstream evaluation on all $500$ SWE-Bench Verified instances
completes with the same wall-clock budget as the CodeGrep downstream
evaluations described in \S\ref{sec:setup}.

\section{Case Study: Action-Level Trajectory}
\label{app:case-trajectory}

Table~\ref{tab:case-django-trajectory} shows the first $18$ tool
calls of the baseline vs.\ CodeGrep-injected trajectories on
\texttt{django-15278}, illustrating the failure mode discussed in
\S\ref{sec:case}.

\begin{table*}[t]
\centering
\footnotesize
\renewcommand{\arraystretch}{1.15}
\setlength{\tabcolsep}{4pt}
\caption{Action-level trajectory on \texttt{django-15278}. Baseline
spends its first $18$ actions searching for
\texttt{oauth2} (a third-party name in the issue) and reading
tangential files; CodeGrep injection sends the same agent directly to
\texttt{related.py}.}
\label{tab:case-django-trajectory}
\begin{tabular}{@{}r l p{4.2cm}@{\hskip 10pt}r l p{4.2cm}@{}}
\toprule
\multicolumn{3}{c}{\textbf{Baseline (unresolved, $74$ actions)}} &
\multicolumn{3}{c}{\textbf{CodeGrep (resolved, $15$ actions)}} \\
\cmidrule(lr){1-3}\cmidrule(l){4-6}
\# & Tool & Target & \# & Tool & Target \\
\midrule
1--2   & think        & analyze issue                     & 1--5   & think        & analyze + read hints \\
3      & task         & plan                              & 6      & file\_editor & \texttt{related.py} \\
4--9   & terminal     & \texttt{find | grep oauth2}$\times 6$ & 7  & think        & locate OneToOneField \\
10     & think        & rewrite plan                      & 8      & terminal     & \texttt{grep OneToOneField} \\
11--14 & terminal     & \texttt{grep AddField}            & 9      & file\_editor & \texttt{related.py} \\
15     & file\_editor & \texttt{tests/.../test\_state.py} & 10--11 & think        & propose fix \\
16--17 & terminal     & \texttt{grep -n} for line hits    & 12     & file\_editor & edit \texttt{\_\_init\_\_} \\
18     & file\_editor & \texttt{test\_state.py} l.\,$627$ & 13     & think        & verify \\
19--20 & think        & propose migration-op fix          & 14     & finish       & --- \\
21     & file\_editor & \texttt{operations/fields.py}     & \multicolumn{3}{c}{} \\
$\cdots$ & $\cdots$   & (never converges)                 & \multicolumn{3}{c}{} \\
\bottomrule
\end{tabular}
\end{table*}


\end{document}